\documentclass[twoside,twocolumn,9pt]{article}
\usepackage{extsizes}
\usepackage[super,sort&compress,comma]{natbib} 
\usepackage[version=3]{mhchem}
\usepackage[left=1.5cm, right=1.5cm, top=1.785cm, bottom=2.0cm]{geometry}
\usepackage{balance}
\usepackage{times,mathptmx}
\usepackage{sectsty}
\usepackage{graphicx} 
\usepackage{lastpage}
\usepackage[format=plain,justification=justified,singlelinecheck=false,font={stretch=1.125,small,sf},labelfont=bf,labelsep=space]{caption}
\usepackage{float}
\usepackage{fancyhdr}
\usepackage{fnpos}
\usepackage[english]{babel}
\addto{\captionsenglish}{%
  
}
\usepackage{array}
\usepackage{droidsans}
\usepackage{charter}
\usepackage[T1]{fontenc}
\usepackage[usenames,dvipsnames]{xcolor}
\usepackage{setspace}
\usepackage[compact]{titlesec}
\usepackage{hyperref}
\usepackage{bm}

\usepackage{subcaption}

\definecolor{cream}{RGB}{222,217,201}

\begin{document}

\pagestyle{fancy}
\thispagestyle{plain}
\fancypagestyle{plain}{

\fancyhead[C]{}
\fancyhead[L]{}
\fancyhead[R]{}
\renewcommand{\headrulewidth}{0pt}
}

\makeFNbottom
\makeatletter
\renewcommand\LARGE{\@setfontsize\LARGE{15pt}{17}}
\renewcommand\Large{\@setfontsize\Large{12pt}{14}}
\renewcommand\large{\@setfontsize\large{10pt}{12}}
\renewcommand\footnotesize{\@setfontsize\footnotesize{7pt}{10}}
\makeatother

\renewcommand{\thefootnote}{\fnsymbol{footnote}}
\renewcommand\footnoterule{\vspace*{1pt}%
\color{cream}\hrule width 3.5in height 0.4pt \color{black}\vspace*{5pt}} 
\setcounter{secnumdepth}{5}

\makeatletter 
\renewcommand\@biblabel[1]{#1}            
\renewcommand\@makefntext[1]%
{\noindent\makebox[0pt][r]{\@thefnmark\,}#1}
\makeatother 
\renewcommand{\figurename}{\small{Fig.}~}
\sectionfont{\sffamily\Large}
\subsectionfont{\normalsize}
\subsubsectionfont{\bf}
\setstretch{1.125} 
\setlength{\skip\footins}{0.8cm}
\setlength{\footnotesep}{0.25cm}
\setlength{\jot}{10pt}
\titlespacing*{\section}{0pt}{4pt}{4pt}
\titlespacing*{\subsection}{0pt}{15pt}{1pt}

\fancyfoot{}
\fancyfoot[LO,RE]{}
\fancyfoot[CO]{}
\fancyfoot[CE]{}
\fancyfoot[RO]{\footnotesize{\sffamily{1--\pageref{LastPage} ~\textbar  \hspace{2pt}\thepage}}}
\fancyfoot[LE]{\footnotesize{\sffamily{\thepage~\textbar\hspace{2pt} 1--\pageref{LastPage}}}}
\fancyhead{}
\renewcommand{\headrulewidth}{0pt} 
\renewcommand{\footrulewidth}{0pt}
\setlength{\arrayrulewidth}{1pt}
\setlength{\columnsep}{6.5mm}
\setlength\bibsep{1pt}

\makeatletter 
\newlength{\figrulesep} 
\setlength{\figrulesep}{0.5\textfloatsep} 

\newcommand{\topfigrule}{\vspace*{-1pt}%
\noindent{\color{cream}\rule[-\figrulesep]{\columnwidth}{1.5pt}} }

\newcommand{\botfigrule}{\vspace*{-2pt}%
\noindent{\color{cream}\rule[\figrulesep]{\columnwidth}{1.5pt}} }

\newcommand{\dblfigrule}{\vspace*{-1pt}%
\noindent{\color{cream}\rule[-\figrulesep]{\textwidth}{1.5pt}} }

\makeatother

\twocolumn[
  \begin{@twocolumnfalse}
\vspace{3cm}
\sffamily
\begin{tabular}{m{4.5cm} p{13.5cm} }

 & \noindent\LARGE{\textbf{Marangoni stress induced by rotation frustration in a liquid foam $^\dag$}} \\
\vspace{0.3cm} & \vspace{0.3cm} \\

 & \noindent\large{Antoine B\'erut and  Isabelle Cantat } \\

 & \noindent\normalsize{The role of surface tension gradients in the apparent viscosity of liquid foams remains largely unexplained. In this article, we develop a toy-model based on a  periodic array of 2D hexagonal bubbles, each bubble being separated from its neighbors by a liquid film of uniform thickness. The two interfaces of this thin liquid film are  allowed to slide relatively to each other, thus shearing the liquid phase in between.
We solve the dynamics under external shear of this minimal system and we show that the continuity of the surface tension around the whole bubble is the relevant condition to determine the bubble rotation rate and the energy dissipation. This result is expected to be robust in more complex situations and illustrates that thin film dynamics should be solve  at the scale of the whole bubble interface when interface rheology matters. 
 } \\

\end{tabular}

 \end{@twocolumnfalse} \vspace{0.6cm}

  ]

\renewcommand*\rmdefault{bch}\normalfont\upshape
\rmfamily
\section*{}
\vspace{-1cm}


\footnotetext{Univ Rennes, CNRS, IPR (Institut de Physique de Rennes) - UMR 6251, F- 35000 Rennes ; E-mail: isabelle.cantat@univ-rennes1.fr}

\footnotetext{\dag~Electronic Supplementary Information (ESI) available: The Matlab code that generates the data presented in this article is available in Zenodo repository: \href{https://doi.org/10.5281/zenodo.1409638}{10.5281/zenodo.1409638}.}


\newcommand{\ep}{\dot{\varepsilon}^{\mathrm{ext}}}
\section{Introduction}
\label{intro}

The rheological properties of foams are crucial for most industrial applications involving flowing or deforming foams. 
However, the flows induced in the liquid phase, at the bubble scale, during the deformation of a foam sample has never been fully characterized. This explains why the effective foam viscosity remains difficult to predict, as a function of the physical parameters of the foam and of the  physico-chemical parameters of the foaming solution \cite{livre_mousse_en, cohenaddad13, dollet14b}.  
The liquid phase is a network of thin liquid films and thicker menisci, also called Plateau borders. The various possible flows in this network have been  well identified by Buzza {\it et al.} \cite{buzza95} but their relative importance remains unclear. However, as the viscous dissipation is an increasing function of the confinement, the effective viscosity of the foam is expected to be governed for a large part by the dissipation in the thin films,  where the viscous phase is the most confined. 

Pioneer rheological models of foam described bubbles as soft solid spheres separated by flat films of uniform thickness \cite{durian95, denkov08}, and obtained important results in the field.
However, a fundamental difference between soft solid spheres and bubbles is their ability to impose a pressure gradient in these flat films: a solid can, whereas a bubble can not. More precisely, if a gas bubble is at uniform pressure,  the pressure  in the liquid phase, at the interface, is given  by the Laplace pressure jump, which vanishes for flat films. Moreover, given the quasi-parallel nature of the velocity field in the films, pressure gradients in the direction normal to the interface are negligible.   Consequently, the pressure in a flat film separating two bubbles is uniform. The only driving forces for the flow are thus the surface tension gradient, {\it i. e.} the Marangoni forces. 

Some numerical models have been  able to take into account all the physical ingredients governing the flow in a 2D periodic system  \cite{pozrikidis01, titta18}, or in a disordered one, at the price of more important simplifications \cite{kahara17}.  However, toy models are still of high importance, as they allow to build a simple intuition of the physical processes governing the flows.
We thus revisit the well known Princen's model \cite{princen83} in an out of equilibrium context,  to shed light on the Marangoni forces governing the shear rates in the films, in the simplest possible example.

We consider a 2D periodic foam made of hexagonal bubbles of identical area $\mathcal{A}$. We assume that bubbles are separated by a film of uniform and constant thickness $h$ much smaller than the bubble size ($h\ll\sqrt{\mathcal{A}}$), and that  the classical Plateau equilibrium rules  remain valid during the deformation: films are represented by straight edges and  three edges meet at 120$^\circ$ at menisci of negligible size (see Fig. \ref{notation}).
With these crude assumptions, the normal velocities of the film interfaces induced by an imposed shear deformation of the foam is well known \cite{princen83}. The aim of our model is to establish rules for  the {\it tangential} velocities of these interfaces, on both sides of the films.  Tangential  velocities are directly related to the thin films shearing, and thus to the viscous dissipation. They  are therefore the most relevant quantities needed to build an effective foam viscosity. 
In the limit of quasi-inextensible interfaces addressed in this article, a single degree of freedom remains on each bubble:  the rate at which the interface rotates around the bubble shape, when the foam is sheared.
For a given film separating two bubbles, it is always possible to chose the rotation rate of the interface on these two bubbles so that the film is not sheared. However, a bubble is in contact with 6 films, as it has 6 neighboring bubbles,  and no rotation rate can insure a vanishing shear in each of  these six films. A geometric frustration emerges, in a similar spirit as the rolling frustration introduced in the context of granular flows \cite{herrmann90}, and the rotation rate of the bubbles arises from a global optimization of the system that we make explicit for the simple case of a periodic array of bubbles.

In that case, we show that a unique rotation rate satisfies the following physical constraint: the surface tension must remains continuous around the whole bubble. This seemingly obvious property actually imposes a non local constraint on the surface tension gradient, {\it ie} on the Marangoni forces: the integral of the Marangoni force around a bubble vanishes. 
We determine the interface rotation rate from this global constraint, as well a the induced surface tension variations around the bubble,  as a function of the bubble elongation and of its  orientation with respect to the shear.  This approach  could be extended to more complex situations, with more realistic interface rheological properties, or with variable film thicknesses. An important consequence of the non locality of the constraint, for flows involving high interface elasticity or viscosity, is that the whole interface of the physical system  must be considered, otherwise  the tangential velocity remains undetermined.

\section{Model}
\label{model}

\begin{figure}[!ht]
\centering
\includegraphics[width=0.45\textwidth]{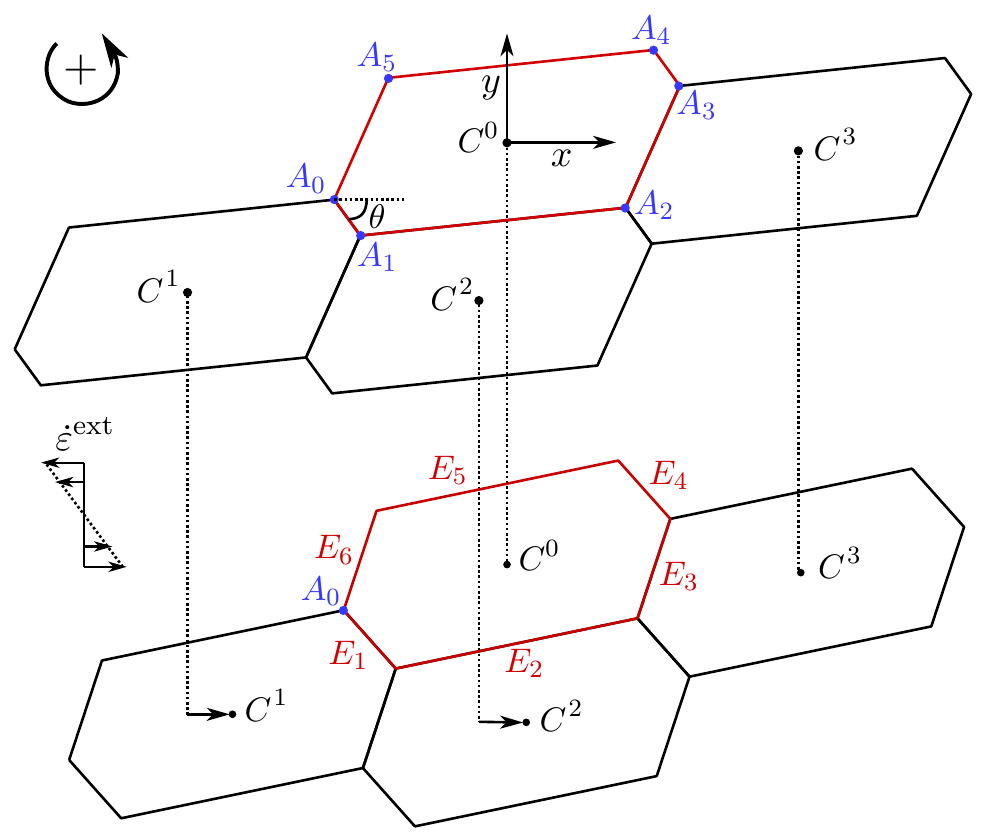}
\caption{Example of the 2D periodic  hexagonal foam at time $t$ (top) and $t+dt$ (bottom), and notations used in the text. In this case, at time $t$, the control parameters are  $L_1/L= \alpha=0.1 $, $(L_1+L_2)/L= \beta=0.7$ and $\theta=-54^\circ$. }\label{notation}
\end{figure}

We consider a 2D periodic foam made of centro-symmetric hexagons of area $\mathcal{A}$.
Our  notations are shown in Fig. \ref{notation} :  we choose a reference hexagon $H^0$ of center $C^0$ in the periodic structure and we denote its  vertices by $A_i$, with $i \in [0 - 5]$ ; the segment $[A_{i-1} A_i]$ is the edge $E_i$ of length $L_i$ ; the hexagon perimeter is $2L$, with $L=L_1+ L_2+ L_3$ ; the edge $E_1$ makes an angle $\theta$ with the direction $x$ (counted positively in the anti-clock wise orientation); the hexagon sharing the edge $E_i$ with the reference hexagon is denoted $H^i$, of center $C^i$.   At each vertex, and at all times, we assume that the angles between the edges are  $120^\circ$  as imposed by the equilibrium Plateau rules \cite{livre_mousse_en}.

The vertex  $A_{i+3}$ is the symmetric of the vertex  $A_i$ with respect to $C^0$, and the foam dynamics thus only needs to be solved along the edges $E_1$, $E_2$ and $E_3$. 
The position along these edges is measured by the curvilinear abscissa $S$, with the reference $S=0$ at the vertex $A_0$ (at all time). However, the main spatial variable 
of the model is the non-dimensional curvilinear abscissa $s=S/L$.  The abscissa of the vertices  are, by definition,  $s(A_0)=0$, $s(A_1)=L_1/L=\alpha$, $s(A_2)=(L_1+L_2)/L=\beta$ and $s(A_3)=1$. 
There is no inertia in our model, so the evolution of the foam between two times $t$ and $t+dt$ does not depend on its history, but only on its shape at the time of interest, entirely determined by the three control  parameters   $\alpha$,  $\beta$ and $\theta$. 

A simple shear of rate  $\ep$ is imposed to the foam.   In a complex fluid, as a foam,  the external shear only controls the large scale deformation of the structure. If this structure is periodic, the external shear actually acts on the position of the periodic cells, {\it i.e.} in our case, on the position of the bubble centers. 
Between the times $t$ and $t+dt$, the center $x_C,y_C$ of each hexagon  thus moves with the rule
\begin{equation}
x_C(t+dt) = x_C(t) - \ep\,  y_C(t) \, dt  \quad; \quad   y_c(t+dt)=y_C(t) \; .
\label{ext_shear} 
\end{equation} 
Note that with this sign convention, a positive value of $\ep$ induces a positive  local rotation rate (see Fig. \ref{notation}).  

In contrast, inside a periodic cell, the local structure follows a non affine motion. For fast deformations, out of equilibrium angles are expected between the edges \cite{cantat11, grassia12}. Here we assume that the equilibrium rule for the angles remains valid under shear and we impose that each  vertex  moves in order to keep an angle of $120^\circ$ between the edges.  Note that in a more refined model, it could be replaced by any other  rule, without modification of  the remaining part of the modelisation.
In this theoretical  frame, the vertex position is given by a unique function of the position of the three adjacent bubble centers (given latter in eq. \ref{pos_vertex}). The foam structure at the time $t+dt$ is thus a complex, but explicit, function of the various control parameters at time $t$. Especially, the new values of the shape parameters $\alpha(t+dt)=\alpha+d\alpha$ and  $\beta(t+dt)=\beta+d\beta$ can be expressed as a function of $\alpha(t)$, $\beta(t)$, $\theta(t)$ and $\ep$ .
If one edge is too short at time $t$, it may happen that no equilibrium shape exists at $t+dt$ unless a bubble rearrangement $T1$ occurs \cite{livre_mousse_en}. We will not consider this case in the following.

The foam structure evolution under shear described above is simply the one of the classical Princen's model \cite{princen83}. However, the novelty  is to consider, in the simplest possible way, the consequences of this given  structure deformation on the relative motion of the foam film interfaces, and thus, on the internal viscous dissipation and  Marangoni stress.

\begin{figure}[!ht]
\centering
\includegraphics[width=0.45\textwidth]{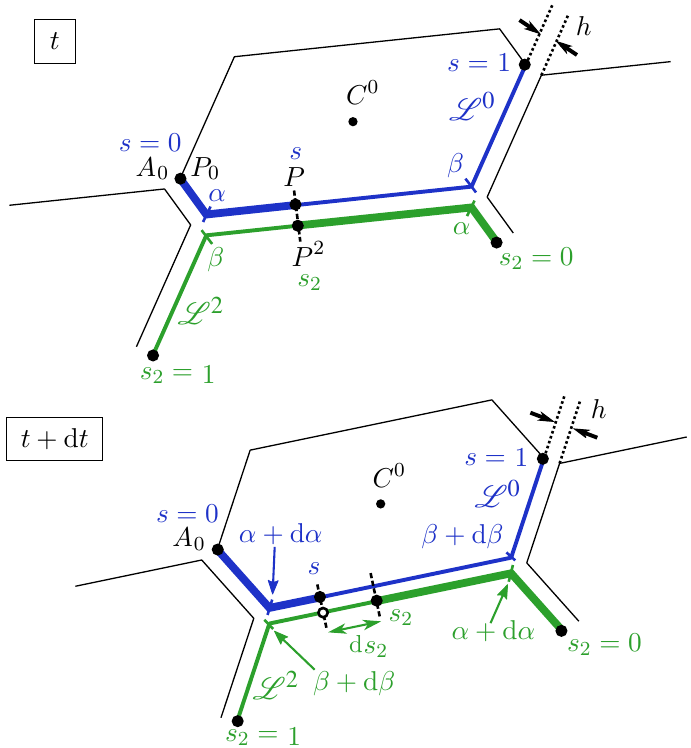}
\caption{Illustration of the interfacial motion during step (I), with increased value of $h$ for readability reasons. The material system ${\mathcal{L}}^0$ (dark blue online) and its symmetric periodic image ${\mathcal{L}}^2$ (green online) are shown at time $t$ (top) and $t+dt$ (bottom). Material points belonging to these different systems are represented by the symbols $\bullet$, and their rescaled abscissa are given.  
By convention, the origin of $s$  is at the vertex $A_0$ on ${\mathcal{L}}^0$ and the origin of $s_2$ is at the symmetric image of $A_0$ on  ${\mathcal{L}}^2$. 
The material point of interest is $P$, belonging to ${\mathcal{L}}^0$ and at the position $s$ at $t$. At that time, its coincident point on ${\mathcal{L}}^2$ is $P^2$ at $s_2$.  
The thick line on ${\mathcal{L}}^0$ is the material system bounded by $P_0$ (with $s(P_0,t)=0$) and $P$, followed between times $t$ and $t+dt$. 
Similarly, on ${\mathcal{L}}^2$, the thick line is the material system between the point verifying $s_2(t)=0$ and  $P^2$.
As imposed during step (I), the initial point $P_0$ stay at the vertex $A_0$  and thus at abscissa $s=0$, as well as its image on ${\mathcal{L}}^2$.
As the rescaled length of both material systems remains constant, the abscissa of $P$ and $P^2$ are still $s$ and $s_2$ at $t+dt$.
However $P^2$ is not in front of $P$ anymore. It has been replaced by a new point  $P^2(t+dt)$, at abscissa $s_2+ds_2$, represented by ${\mathbf{\circ}}$.
}\label{stepA}
\end{figure}

The gas bubble $H^0$ is covered by a continuous surfactant monolayer of length $2 L$. The part of this layer located along the edges $E_1$, $E_2$ and $E_3$ at time $t$ is called ${\mathcal{L}}^0$: using a continuous medium approach, we consider ${\mathcal{L}}^0$ as a material system in which each point can be followed along its trajectory. 
One side of the liquid film represented by the edge $E_i$ is thus covered by ${\mathcal{L}}^0$. Its other side is covered by the material system  called ${\mathcal{L}}^i$, which is a symmetric image of ${\mathcal{L}}^0$, as depicted in Fig. \ref{stepA}.  
Disregarding the specific role of the Plateau borders located at each vertex, we assume that the liquid film confined between these two  interfaces has a constant and uniform thickness $h$. This thickness is much smaller than the bubble size ($h\ll\sqrt{\mathcal{A}}$) and a material point of interface on the edge $E_i$ will be assumed to be at the same location $x,y$ whether it belongs to the interface ${\mathcal{L}}^i$ or to the interface  ${\mathcal{L}}^0$.
However, any relative tangential velocity $\delta v$  between the two facing interfaces induces a shear flow in the thin liquid film of viscosity $\eta$, and thus a viscous stress $\eta \delta v/h$ in which the finite value of the thickness $h$ is taken into account.   

To build the simplest possible model, we further  assume that  the small  variation $\Delta L$ of the bubble perimeter $L$, imposed by the global shear deformation between times $t$ and $t+dt$, induces a compression or stretching of the interface which is homogeneous over the whole layer ${\mathcal{L}}^0$. We show in Appendix that this happens when the interface has a very large Gibbs modulus $E$. In that limit, the mean value of the surface tension increases with time proportionally to $E$ and to $\Delta L$, but the surface tension gradients do not depend on  $E$. As we only focus on the spatial variation of the tension, and not on its temporal evolution, the Gibbs modulus thus plays no role in the problem. The validity range of this large $E$ assumption is discussed in the section~\ref{results}.  

Using this assumption, the rescaled distance $s(P)-s(P_0)$ between two material points $P$ and $P_0$ in  ${\mathcal{L}}^0$  remains constant during the deformation. Consequently, the rescaled position at $t +dt$ of all points in ${\mathcal{L}}^0$ is fully determined by the position of any material point $P_0$ in ${\mathcal{L}}^0$. In the following we use arbitrarily as reference point $P_0$ the point located at $A_0$ at time $t$,  {\it i.e.} verifying $s(P_0, t)=0$. 
The  rescaled abscissa of $P_0$ at the time $t+dt$, measuring the rotation of the interface over itself,  is the only degree of freedom we keep in our toy model, and for which an equation of evolution is established in the next paragraph.
We define $s^*\equiv u^* dt=s(P_0, t+dt)$. If $s^*>0$, the point $P_0$ is on the edge $E_1$ at $t+dt$ and if $s^*<0$, it is on the edge $E_6$. Note that $A_0$ is taken as a fixed reference point: $A_0$ is not a material point and verifies $s(A_0,t)=0$ at all times.

For sake of clarity, the interface motion between time $t$ and $t+dt$ will be arbitrarily decomposed in two steps: Step (I), the bubble shape evolves,  the point $P_0$ is maintained at the position $A_0$ (meaning that $s^*=0$ is imposed) ; Step (II), the bubble shape is maintained at its $t+dt$ value and $s^*$ is computed on the basis of the results of step (I).  Step (I) thus addresses the imposed shear deformation, and step (II) the global rotation of the bubble on itself.
By linearity, the viscous stress induced in the liquid films by the step (II) simply adds to the one obtained in step (I). Therefore, this arbitrary decomposition of the motion does not introduce any additional  approximation.

First, we discuss the motion during the step (I), where $P_0$ is fixed at the position $A_0$ and only the shape of the bubble is modified. Let $P$ be a point of abscissa $s$ in ${\mathcal{L}}^0$, on $E_i$. At the time $t$, it is at the same position $x(t),y(t)$ than a point $P^i(t)$, called its coincident point, belonging to ${\mathcal{L}}^i$, on the other side of the thin film $E_i$. Despite the fact that $P$ and $P^i(t)$ are at the same spatial position, the abscissa $s_i(s,t)$, computed on ${\cal L}^i$, differs from $s$, as depicted in Fig. \ref{stepA}, for the case $i=2$. The value of $s_i(s,t)$ can be expressed as a simple function of $s$, $\alpha(t)$ and $\beta(t)$  established  in the section \ref{resolution} (eq. \ref {si}), from the symmetry and periodicity rules. At time $t+dt$, the point $P$ is still at the abscissa $s$ (because $s^*=0$ in step (I)), but at a new position  $x(t+dt),y(t+dt)$. For simplicity, we assume that $P$ has been chosen far away from the vertices to be still on the same edge $E_i$ at $t+dt$. A new material point is in front of it: a point $P^i(t+dt)$, having the abscissa $s_i(s,t+dt) \equiv s_i(s,t) + ds_i $. The two points $P$ and $P^i(t)$ thus moved from a distance $ds_i$ relatively to each other during $dt$: this is the signature of a local shear rate in the thin film of amplitude $\dot{\varepsilon}= ( L ds_i/dt) / h$. As the rescaled distance between two points on the same layer is kept constant, the quantity $ds_i$ is the same for all point $P$ chosen on the same edge $E_i$. It is not defined for the points $P$ that goes from one edge to the other during the time interval $dt$. However these points lead to a second order contribution, that tends to zero at small $dt$ and that can be safely neglected. 

The tangential stress balance at the interface, also called the Marangoni law, imposes that the surface tension gradient $d\gamma/(Lds)$ balances the viscous stress $\eta  \dot{\varepsilon}$. With the orientation conventions we use, we get, on each edge $E_i$ and for the step (I): 
  
\begin{equation}
\frac{1}{L} \left .\frac{d\gamma}{ds} \right )_\mathrm{I} = - \frac{\eta L}{h} \frac{ds_i}{dt} \; .
\label{dgamma_a}
\end{equation}

The surface tension difference along the half bubble perimeter induced by the step (I) is thus 

\begin{equation}
 \gamma^{\; \mathrm{I}}(1)-\gamma^{\; \mathrm{I}}(0) = - \frac{\eta L}{h}\left(L_1 \frac{ds_1}{dt} + L_2 \frac{ds_2}{dt}+ L_3 \frac{ds_3}{dt} \right)\; .
\label{Dgamma_a}
\end{equation}
  
Points $A_0$ and $A_3$ are periodic images of each other, and, by continuity of the surface tension, they must have the same surface tension value, thus imposing $\gamma(1)=\gamma(0)$. This condition will be fulfilled thanks to the additional viscous stress induced by step (II) {\it i.e.} by the rotation of the interfacial layer around each bubble, governed by $s^*$. As shown below, this determines a unique value for the sought parameter $s^*$.

\begin{figure}[!ht]
\centering
\includegraphics[width=0.45\textwidth]{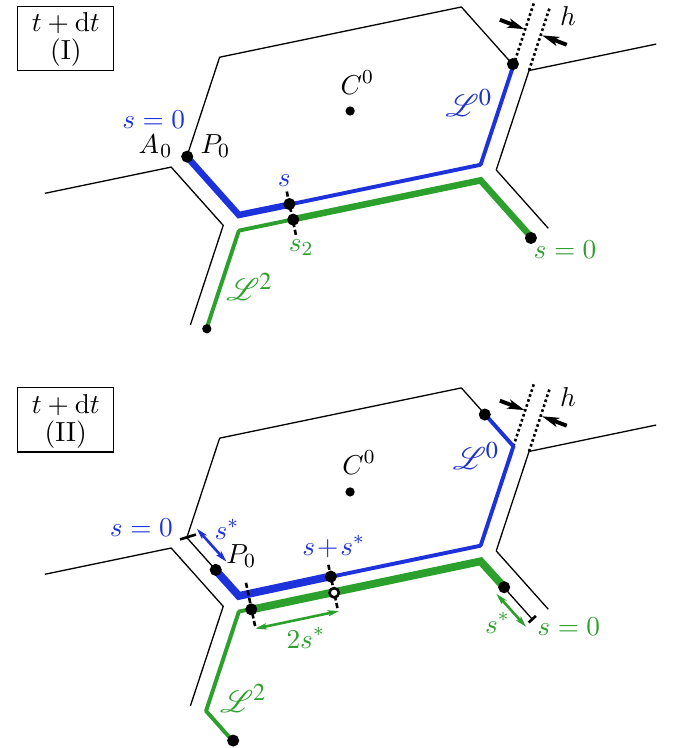}
\caption{Illustration of the interface motion during step (II), with conventions similar as the one in Fig. \ref{stepA}. 
At the end of step (I), a point $P_2$ at abscissa $s_2$ is in front of $P$ at abscissa $s$. Then the reference point $P_0$ moves a distance $s^*$ along the interface, and so does its symmetric image on on ${\mathcal{L}}^2$. After step (II) the distance between $P$ and $P^2$ is $2 s^*$.
}\label{stepB}
\end{figure} 

During step (II), the shape of the bubble is fixed, and the materials points only moves along the perimeter. The point $P$ moves over the distance $s^*= u^* dt$, whereas its coincident point moves over the distance $-s^*$ as shown in Fig. \ref{stepB}. The local shear on all edges is thus $2 L u^*/h$, the surface tension gradient is

\begin{equation}
\frac{1}{L} \left .\frac{d\gamma}{ds} \right )_\mathrm{II} =  2 \frac{\eta L}{h} u^* \; , 
\label{dgamma_b}
\end{equation}

and the surface tension difference induced by step (II) is  

\begin{equation}
\gamma^{\; \mathrm{II}}(1)-\gamma^{\; \mathrm{II}}(0) =  \frac{2 \eta L^2 }{h} u^* \; .
\label{Dgamma_b}
\end{equation}

The condition  $\gamma^{\; \mathrm{I}}(1)+\gamma^{\; \mathrm{II}}(1)= \gamma^{\; \mathrm{I}}(0)+\gamma^{\; \mathrm{II}}(0)$ then provides the expression for $u^*$:
\begin{equation}
 u^* = \frac{1}{2 L}\left(L_1 \frac{ds_1}{dt} + L_2 \frac{ds_2}{dt}+ L_3 \frac{ds_3}{dt}\right) \; .
\label{ustar}
\end{equation}
   
The determination of this rotational velocity gives access to the dynamical quantities of interest : the viscous dissipation and the relative amplitude of the surface tension fluctuations. 

Finally, note that both $\left . \frac{d\gamma}{ds} \right)_\mathrm{I}$, $\left . \frac{d\gamma}{ds} \right)_\mathrm{II}$, and $u^*$ depend on the arbitrary choice of $A_0$ as the reference point for $s=0$. However, the actual  motion of the material points, resulting from the sum of step (I) and step (II), does not depends on this choice, and identical values are obtained for any other fixed point.
In the next section, we build the explicit relationships between the initial bubble shape, the imposed shear,  and the various physical quantities introduced in this section. 

\section{Analytical resolution}
\label{resolution}

\begin{figure}[!ht]
\centering
\includegraphics[width=0.45\textwidth]{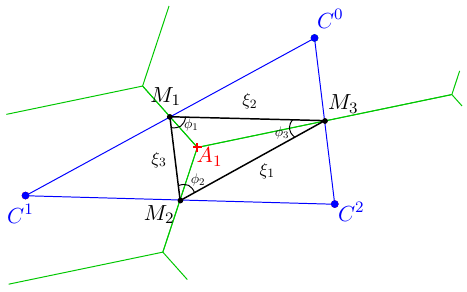}
\caption{Determination of the position of $A_1$ at time $t+dt$.  The points $C^0$, $C^1$ and $C^2$ are the bubble centers at time $t+dt$ and the points  $M_1$, $M_2$ and $M_3$ are the middles of the edges of the triangle $C^0 C^1 C^2$. The edge lengths $\xi_i$ and angles $\phi_i$ of the triangle $M_1 M_2 M_3$ are the quantities involved in the equation \ref{pos_vertex}, which allows to determine the position of  $A_1$ at time $t+dt$.}\label{fermat}
\end{figure}

The model discussed in the previous section can be analytically solved for any set of the control parameters
  $\alpha$, $\beta$ and  $\theta$, which  characterize the initial shape of the hexagon and its orientation with respect to the shear. For each set of values $(\alpha,\beta, \theta)$, we first determine the foam geometry at time $t$: the half perimeter $L$ of the corresponding hexagon of area $\mathcal{A}$, and the center position of the different hexagons in the network. Using eq. \ref{ext_shear}, we then  determine the positions  $C^0$, $C^1$ and $C^2$ of the centers at $t+dt$. 
To compute the non affine motion of the vertices,  we define the points $M_1$, $M_2$ and $M_3$, respectively the middle of the segments $[C^0 C^1]$,  $[C^1 C^2]$ and $[C^2 C^0]$ (see Fig. \ref{fermat}).
The point $A_1$ that insures angles of $120^\circ$ at the vertex at $t+dt$  is the Fermat point of the triangle $M_1 \, M_2 \, M_3$, given by 
\begin{equation}
 \overrightarrow{OA}_1= \frac{a_1 \xi_1}{K}  \overrightarrow{OM}_1 + \frac{a_2 \xi_2}{K}  \overrightarrow{OM}_2 + \frac{a_3 \xi_3}{K}  \overrightarrow{OM}_3
\label{pos_vertex}
\end{equation}
Where $a_i=1/\sin(\phi_i + \pi/3)$ and $K=a_1 \xi_1+a_2 \xi_2+a_3 \xi_3$.  In the triangle $M_0 \, M_1 \, M_2$,  $\phi_i$ is the angle at the vertex $M_i$ and $\xi_i$ is the length of the triangle edge opposite to the vertex $M_i$ (see Fig. \ref{fermat}).

Then, the positions of the other vertices are deduced from the positions of $A_1$, $M_1$, $M_2$  and $M_3$,   using $\overrightarrow{A_1 A}_0 = 2\overrightarrow{A_1 M}_1$ , $\overrightarrow{A_1 A}_2 = 2\overrightarrow{A_1 M}_3$,  and $\overrightarrow{A_2 A}_3= - 2 \overrightarrow{A_1 M}_2$. 

These expressions eventually provide a complex but  explicit expression of $L(t+dt)= |A_0 A_1| +  |A_1 A_2| + |A_2 A_3|$, $\alpha(t+dt)= |A_0 A_1| /L(t+dt) $,  and $\beta(t+dt) = \alpha(t+dt)  +   |A_1 A_2| /L(t+dt) $ as a function of the initial values $\alpha$, $\beta$ and $\theta$. 

\begin{figure}[!ht]
\centering
\includegraphics[width=0.4\textwidth]{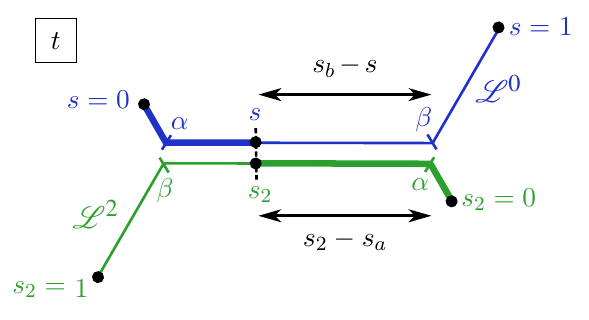}
\caption{Zoom on the liquid film corresponding to the edge $E_2$ of Fig. \ref{stepA}. The point $P$ on ${\mathcal{L}}^0$ is characterized by the abscissa $s$. Its coincident point $P^2$ on ${\mathcal{L}}^2$ is characterized by the abscissa $s_2$. The distance between $P$ and the vertex at $s_b=\beta$ on ${\mathcal{L}}^0$ is the same as the distance between $P^2$ and the vertex at $s_a=\alpha$ on ${\mathcal{L}}^2$, leading to equation~\ref{si}.} \label{relation_ss2}
\end{figure}

At this stage, the normal motion of the foam is known, and the tangential motion of the interface can be computed.
In order to determine the  lengths $ds_i$ introduced in Fig. \ref{stepA}, let consider a point $P$ of ${\mathcal{L}}^0$, of  abscissa $s$, on the edge $E_i$. We call $s_a$ and $s_b$ the abscissa, computed on ${\mathcal{L}}^0$,  of the two vertices $A_{i-1}$ and $A_i$ bounding $E_i$, with $s_a<s<s_b$. The Fig. \ref{relation_ss2} illustrates the case $i=2$ for which $s_a(t)= \alpha(t)$ and $s_b(t)=\beta(t)$. The coincident point $P^i$ is by definition at the same distance of $A_i$ than $P$. However $P^i$ belongs to ${\mathcal{L}}^i$ and, on this layer, by symmetry, the abscissa of $A_i$ is $s_a$ and the abscissa of $A_{i-1}$ is $s_b$. We thus get the condition 
\begin{equation}
s_i(t)- s_a(t) =  s_b(t)-s(t) \, .
\label{si}
\end{equation}

Using eq. \ref{si} at time $t+dt$, we get $s_i(t+dt) = s_a(t+dt) + s_b(t+dt)-s(t+dt)$. After step (I) the abscissa of the material points on ${\mathcal{L}}_0$ are unchanged because $s^*=0$ in step (I), so  $s(t+dt) = s(t)$. Using $s(t)= s_b(t) + s_a(t) - s_i(t)$, we finally get:
\begin{equation}
s_i(t+dt) = s_a(t+dt) + s_b(t+dt)  - s = ds_a +ds_b  + s_i(t)
\label{sidt}
\end{equation} 
The relation $ds_i= ds_a +ds_b$, on the different edges, then leads to the formulas:
\begin{equation}
ds_1= d\alpha \quad \, \quad ds_2 = d\alpha + d \beta  \quad \, \quad ds_3 =  d \beta 
\label{dsi}
\end{equation}

Then, from eq. \ref{ustar}, we get 
\begin{equation}
u^*=\frac{1}{2} \left( \alpha \dot{\alpha} + (\beta-\alpha) (\dot{\alpha} + \dot{\beta}) + (1-\beta) \dot{\beta} \right) \, .
\end{equation}
Finally, the surface tension gradient are obtained by summing eq. \ref{dgamma_a} and eq. \ref{dgamma_b}, leading to  
\begin{eqnarray}
\label{eq:dgammads1}
\left . \frac{\partial \gamma}{\partial s} \right)_1&=& \eta \frac{L^2}{h}(2u^* - \dot{\alpha}) \; , \\
\label{eq:dgammads2}
\left .\frac{\partial \gamma}{\partial s} \right)_2&=& \eta \frac{L^2}{h} (2 u^* - \dot{\alpha}-\dot{\beta}  )\; , \\
\label{eq:dgammads3}
\left . \frac{\partial \gamma}{\partial s} \right)_3 &=& \eta \frac{L^2}{h}(2 u^* -\dot{\beta})\; , 
\end{eqnarray}
respectively on edges $E_1$, $E_2$ and $E_3$. 

These expression can be made more symmetric using the notation $\ell_1 = \alpha$, $\ell_2= \beta-\alpha$ and $\ell_3= 1-\beta$, corresponding to the fraction of perimeter of each edge. 
The previous relations then take the more elegant and symmetric form:
 \begin{equation}
\left . \frac{\partial \gamma}{\partial s} \right)_i= \eta \frac{\ep \cal{A}}{h} \;  \frac{L^2}{\cal{A}} \left( \ell_{i+2} \frac{d \ell_{i+1}}{\ep dt}  - \ell_{i+1} \frac{d \ell_{i+2}}{\ep dt}   \right) \; ,
\label{dgamma_sym}
\end{equation}
with the convention that $i+k$ is computed modulo 3. 
This expression clearly underlines that the arbitrary choice of origin for the abscissa used to establish the relations (\ref{eq:dgammads1} - \ref{eq:dgammads3}) does not play any role in the physical quantities, which are expressed here  as a function of physical quantities only. The factor $L^2/\cal{A}$ and the last factor are non-dimensional and only depends on the geometrical control parameters $\alpha$, $\beta$ and $\theta$. The scaling for the surface tension fluctuations is  given by the first factor $\eta \ep {\cal A} / h$.

The  dissipation rate $\cal{P}$ in the system can be easily deduced from eq. \ref{dgamma_sym} using the relation 
\begin{equation}
{\cal P} = \frac{1}{\eta}\Sigma_i \left [  \frac{1}{L} \left. \frac{\partial \gamma}{\partial s} \right)_i \right]^2  h L_i  \, .
\label{energy}
\end{equation}

As our model only predicts surface tension gradients, the surface tensions  are only determined to within a constant. The mean surface tension  $\langle \gamma \rangle$, averaged over the whole bubble, thus remains unknown. We use it as integration constant to express $\gamma(s)$.

\begin{figure}[!ht]
\centering
\includegraphics[width=0.5\textwidth]{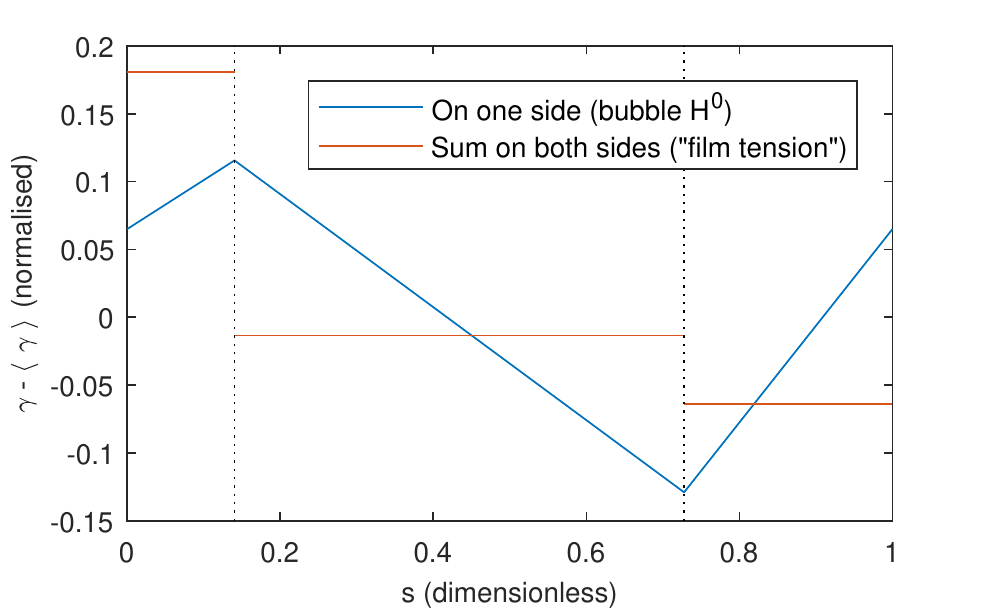}
\caption{Surface tension $\gamma - \langle \gamma \rangle$ normalised by $\eta\mathcal{A}\ep/h$, along the non dimensional curvilinear abscissa $s$, for the initial bubble shape represented in Fig. \ref{notation} ($\alpha=0.1 $, $\beta=0.7$ and $\theta=-54^\circ$ at $t=0$). The surface tension on one side (bubble $H^0$) is shown in blue. The film tension (the sum of the tensions on both sides) is shown in red.}\label{tension}
\end{figure}

In our crude model, the surface tension is a continuous piecewise linear function as shown in Fig.\ref{tension}. 
The  amplitude of its fluctuations can be defined by:
\begin{equation}
\Delta\gamma_{\mathrm{max}} = \max_{[0,L]}\left( \gamma \right) -  \min_{[0,L]}\left( \gamma \right)
\end{equation}
As the tension is a monotonic function on each edge, its extrema are necessarily on the vertices, and $\Delta\gamma_{\mathrm{max}}$ 
is easily determine by comparison of $\gamma(0)$, $\gamma(\alpha)$ and $\gamma(\beta)$. 

We can also define the film tension on each edge, as the sum of the tensions on both sides.  
As shown in Fig. \ref{tension} these film tensions are uniform along a given film and  simply given by $\gamma(0)+\gamma(\alpha)$ on $E_1$, $\gamma(\alpha)+\gamma(\beta)$ on $E_2$, and  $\gamma(\beta)+\gamma(0)$ on $E_3$.
From this we deduce that the maximal difference between two film tensions is equal to $\Delta\gamma_{\mathrm{max}}$.

\section{Results}
\label{results}
We used this model,  implemented in a Matlab code~\cite{matlab_code}, to determine the dynamical evolution of a large set of bubble shapes when an increment of shear $d \varepsilon = \ep dt= 10^{-5}$ is applied.    
The initial shape is characterized by the three parameters $(\alpha, \beta, \theta)$. They have been varied in the range $\{\alpha \in [0.1;0.8]$ and $\beta \in [0.2;0.9]$ such that $\beta - \alpha \geq 0.1\}$ (\textit{i.e.} the rescaled length of each edge is at least $0.1$), and $0^\circ \leq \theta \leq 180^\circ$. As previously stated, the influence of the other physical parameters as $\mathcal{A}$ and $\ep$ are simply deduced from a scaling analysis and do not need to be systematically varied.

\begin{figure}[!ht]
\captionsetup[subfigure]{justification=centering}
\centering
    \begin{subfigure}[b]{0.45\textwidth}
        \includegraphics[width=\textwidth]{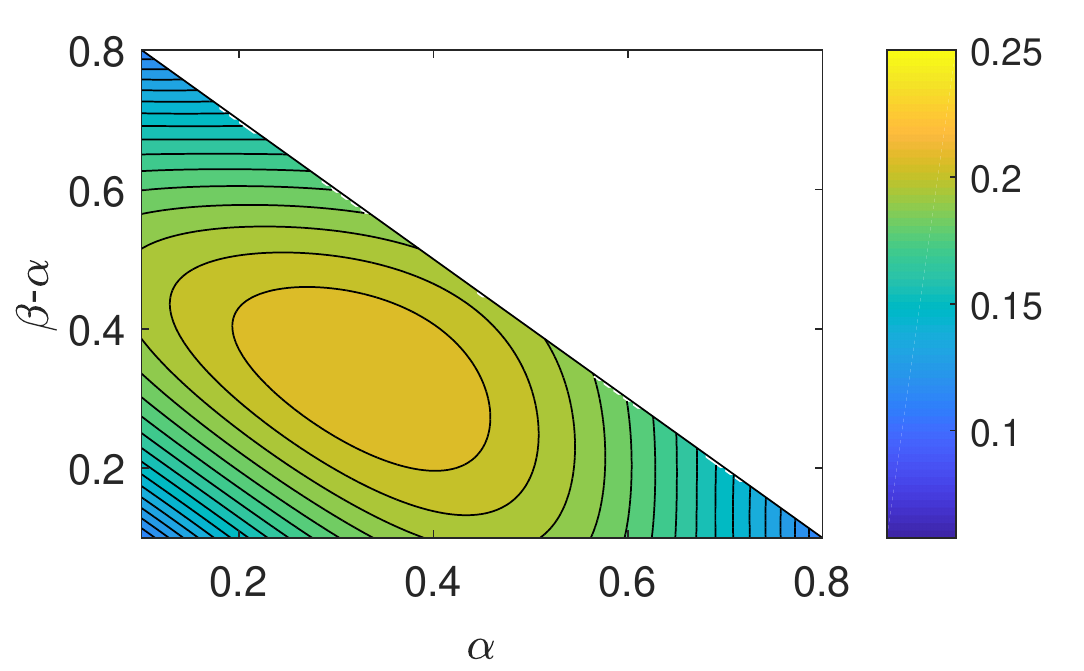}
        \caption{}
        \label{deltagamma_model_a}
    \end{subfigure}
    \begin{subfigure}[b]{0.45\textwidth}
        \includegraphics[width=\textwidth]{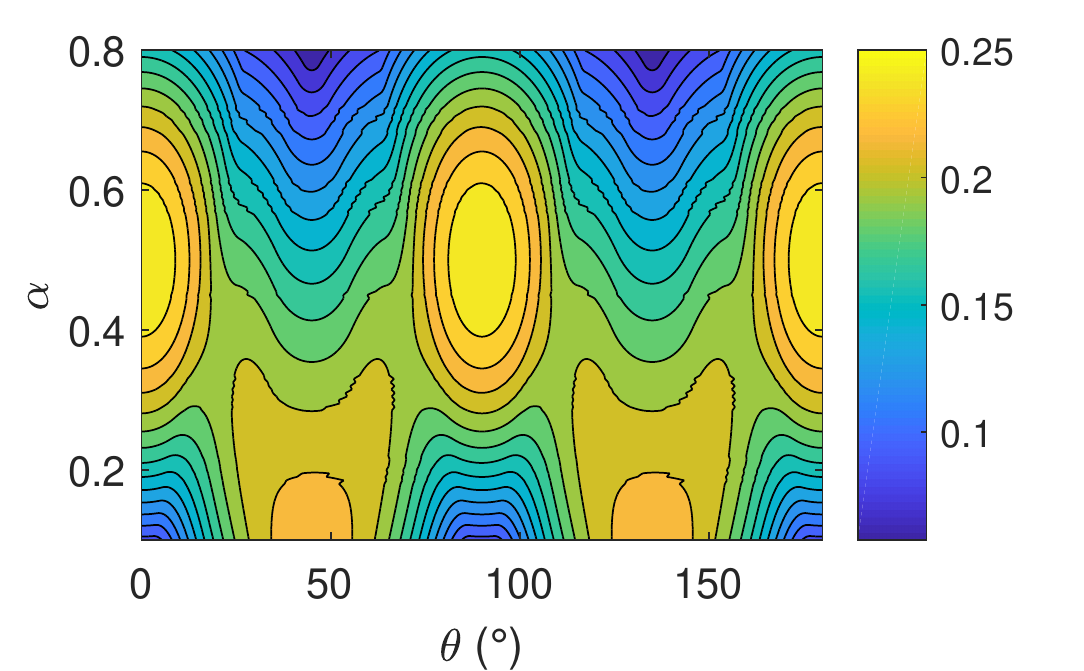}
        \caption{}
        \label{deltagamma_model_b}
    \end{subfigure}
\caption{Color plot of the surface tension fluctuation $\Delta\gamma_{max}$, rescaled by $\eta \mathcal{A} \ep/h$, as a function of the parameters fixing the initial shape of the bubble. \textbf{(a)} $\Delta\gamma_{max}$ averaged over the angle $\theta$, as a function of $\alpha$ and $\beta-\alpha$, the rescaled length of two edges of the hexagon. \textbf{(b)} $\Delta\gamma_{max}$ averaged over the rescaled length $\beta-\alpha$, as a function of $\alpha$ and $\theta$.}\label{deltagamma_model}
\end{figure}

The foam dynamics is first quantified by the value of  $\Delta\gamma_{\mathrm{max}}$. It is represented in Fig. \ref{deltagamma_model_a} as a function of the relative lengths $\alpha$ and $\beta-\alpha$ of the first and second edges of the hexagon. These results are obtained after averaging over the third control parameter $\theta$. Similarly  the influence of $\alpha$ and $\theta$ is shown in Fig. \ref{deltagamma_model_b}, after averaging over $\beta-\alpha$. 

A first result is that  $\Delta\gamma_{\mathrm{max}}$ is of the order of  $(0.15 \pm 0.10) \eta \mathcal{A} \ep/h$ in the whole parameter space. 
Assuming  $\eta=10^{-3}$~Pa.s and $h=10^{-6}$~m, we obtain  $\Delta\gamma_{\mathrm{max}} \approx 0.1$~mN/m  for  $\mathcal{A}=1$~mm$^2$, $\ep=1$~s$^{-1}$. However, for higher shear rates, $\ep=100$~s$^{-1}$ for example, we get  $\Delta\gamma_{\mathrm{max}} \approx 10$~mN/m. Our simple model thus leads to a first conclusion: if $\eta \mathcal{A} \ep/h \ll \gamma$  spatial fluctuations of the surface tension are negligible, otherwise they are not.  
On the basis of this order of magnitude, we can determine the validity range of two key assumptions of the model.  

We first assumed that the angles between films remain equal to 120$^\circ$. This rule arises from the surface tension equilibrium at the vertices and a surface tension difference between adjacent films $\Delta \gamma$ induces an angle modification $\Delta \theta$ scaling as $\Delta \gamma/\gamma$. The geometry imposed in our model thus requires that 
\begin{equation}
\frac{\eta \mathcal{A} \ep}{\gamma h} \ll 1  \; .
\label{contr1}
\end{equation}
A refinement of eq. \ref{pos_vertex} would allow to take into account the out of equilibrium foam shape, as was done in \cite{cantat11}. 

We also assumed  the homogeneity of the extension. As shown in Appendix, this requires that the actual extension $\xi$ can be decomposed into a dominant, uniform, term $\Delta L/L$ and a negligible correction $\delta \xi$, which varies along the bubble contour and has a vanishing mean value.   
For an elastic interface, $\gamma = \gamma_0 + E \xi$, and thus $d \gamma/ds= E d \delta \xi/ds$. The inhomogeneous part of the extension $\delta \xi$ thus remains small as long as $\Delta \gamma \ll E$, leading to a second limitation of the validity range of the model, already discussed in the Appendix :   
\begin{equation}
\frac{\eta \mathcal{A} \ep}{E h} \ll 1  \; .
\label{contr2}
\end{equation}
For foaming solution having a Gibbs elasticity smaller than the surface tension, this second limitation is more restrictive than the one given by eq. \ref{contr1}. 

With this simple scaling argument, we can thus already conclude that models of foam viscosity based on the simple shear of the thin films trapped between bubbles moving at different velocities, as the one developed in this paper,  can only be relevant for small shear rates, small bubbles or thick films. At the opposite limits, surface tension gradients can not be high enough to shear the thin films and another regime should emerge, based on film  extension and compression, and not only on film shearing. 

\begin{figure}[!ht]
\centering
\includegraphics[width=0.25\textwidth]{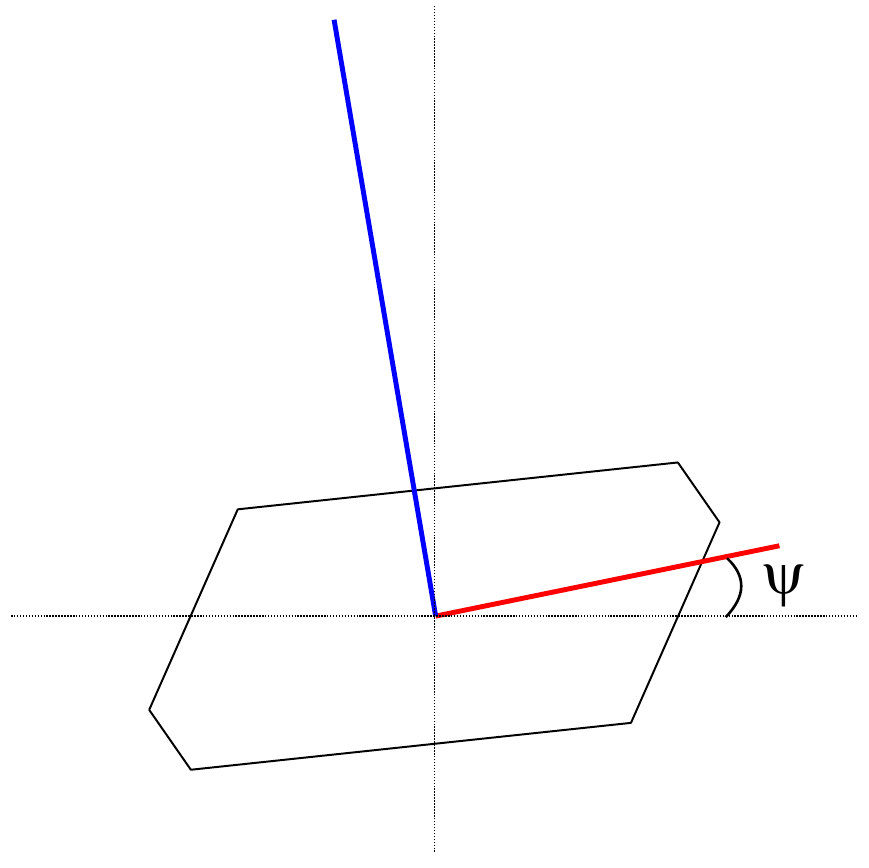}
\caption{Eigenvectors of the 2D inertia matrix of one bubble (the lengths of the red and blue lines are proportional to the square root of the respective eigenvalues). The angle $\psi$ is the angle between the eigenvector associated with the lowest eigenvalue and the horizontal, and the aspect ratio $r$ is the ratio of the square roots of the two eigenvalues. Here $\psi = 16.27^\circ$ and $r=2.47$.}\label{psi_r}
\end{figure}

\begin{figure}[!ht]
\captionsetup[subfigure]{justification=centering}
\centering
    \begin{subfigure}[b]{0.45\textwidth}
        \includegraphics[width=\textwidth]{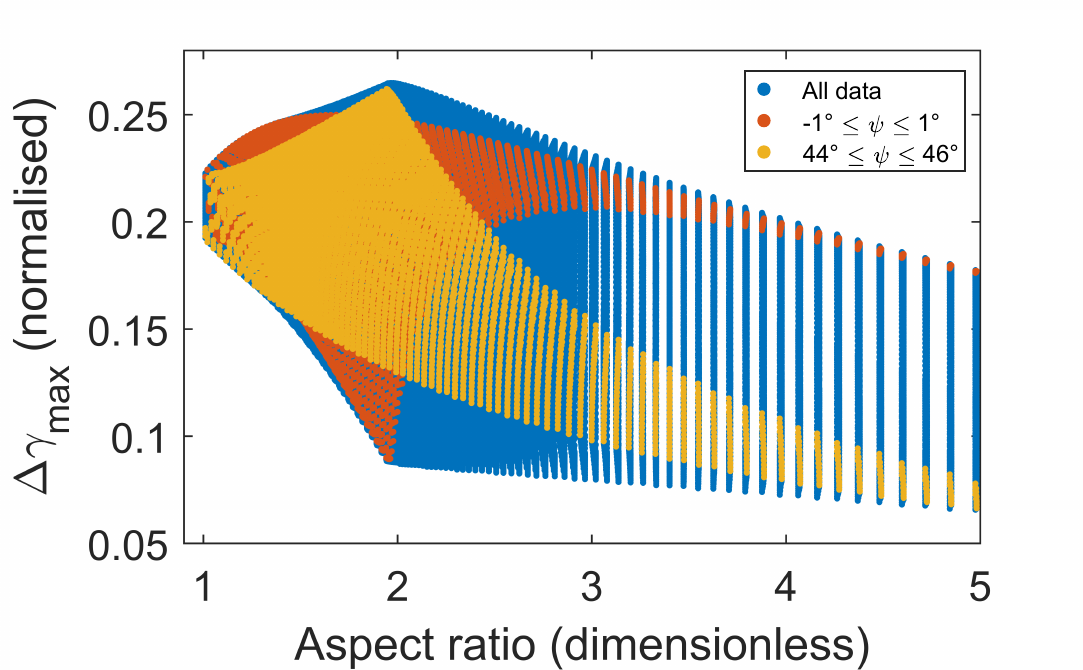}
        \caption{}
        \label{deltagamma_physical_a}
    \end{subfigure}
    \begin{subfigure}[b]{0.45\textwidth}
        \includegraphics[width=\textwidth]{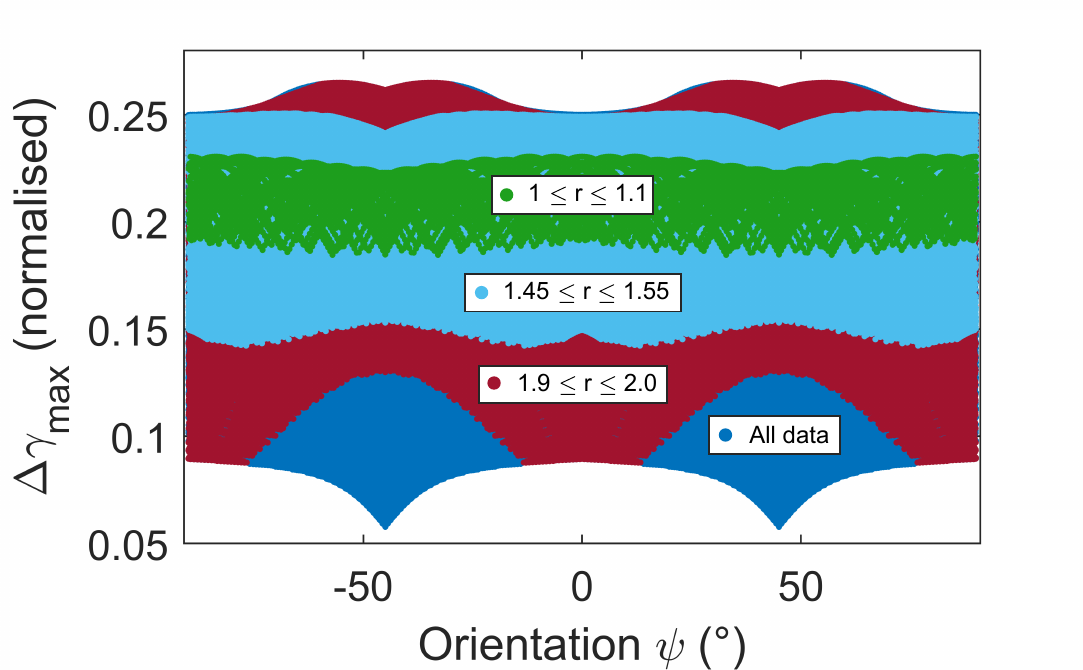}
        \caption{}
        \label{deltagamma_physical_b}
    \end{subfigure}
\caption{Color plot of the surface tension fluctuation $\Delta\gamma_{max}$, rescaled by $\eta \mathcal{A} \ep/h$.  \textbf{(a)} As a function of the aspect ratio of the bubble $r$ (all points in blue), with highlighted regions corresponding to $-1^\circ \leq \psi \leq 1^\circ$ (orange), and $44^\circ \leq \psi \leq 46^\circ$ (yellow). \textbf{(b)} As a function of the orientation of the bubble $\psi$ (all points in blue), with highlighted regions corresponding to $1.9 \leq r \leq 2$ (burgundy),  $1.45 \leq r \leq 1.55$ (light blue), and  $1 \leq r \leq 1.1$ (green).}\label{deltagamma_physical}
\end{figure}

Assuming now that  the constraint eq.\ref{contr2} is fulfilled, we use our model to investigate potential correlations between $\Delta\gamma_{max}$ and the bubble geometry. In this aim, we tried to reduce the complexity to two parameters only, by using the aspect ratio $r$ of the bubble and its  orientation $\psi$ to describe the bubble geometry, instead of using the three parameters $\alpha$, $\beta$ and $\theta$.
These quantities are simply obtained from the 2D inertia matrix of each bubble (taking its center $C^0$ as the origin, and assuming a uniform mass distribution on the edges): $\psi$ is the angle between the eigenvector associated with the lowest eigenvalue and the $x$ direction (horizontal), and $r$ is the ratio of the square roots of the two eigenvalues (see figure~\ref{psi_r}). The values of $\Delta\gamma_{\mathrm{max}}$ as a function of $r$ and $\psi$ are shown in Fig. \ref{deltagamma_physical} (same data as in Fig. \ref{deltagamma_model}).

\begin{figure}[!ht]
\centering
\includegraphics[width=0.5\textwidth]{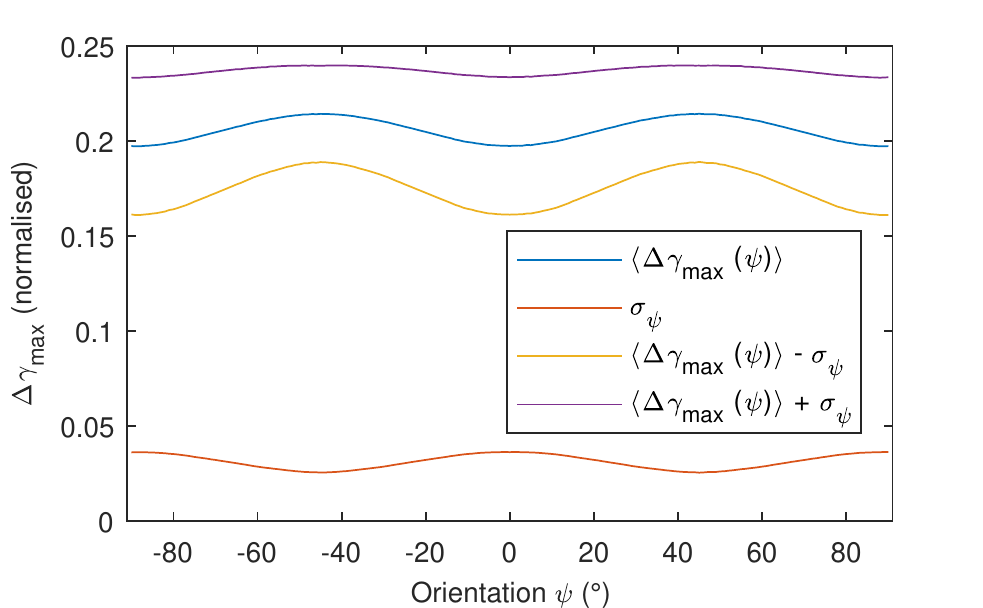}
\caption{Average $\langle \Delta\gamma_{\mathrm{max}} (\psi) \rangle$ and standard deviation $\sigma_\psi$ of surface tension fluctuation on the data subset corresponding to aspect ratio $r\in[1.45;1.55]$ and bubble orientation $[\psi - 0.5^\circ ; \psi+  0.5^\circ]$, as a function of $\psi$.}\label{aspect_ratio_1sur2}
\end{figure}

Different range of aspect ratio may be of interest. In the case of small amplitude oscillatory shear, the average bubble aspect ratio remains close to one, and the surface tension fluctuation  is close to 0.2 $\eta \ep {\cal A} / h$, as shown in Fig. \ref{deltagamma_physical_b}. 
In the case of a steadily sheared foam, a yield strain  is reached, of the order of  $\varepsilon =1$ for disordered 2D foams \cite{weaire94}. It corresponds to a typical bubble aspect ratio of the order of 1.5 \cite{cantat11}. To analyse the surface tension fluctuations in such situation, we focus of the subset of data ${\cal S}$ having $r$ values in the range $[1.45;1.55]$ (corresponding to lightblue points (online) in Fig. \ref{deltagamma_physical_b}). 
In this subset, we get $\langle \Delta\gamma_{\mathrm{max}} \rangle_{\cal S} = (0.206 \pm 0.032) \eta \mathcal{A} \ep/h$, $0.032$ being the  standard deviation $\sigma_{\cal S}$. 
Surprisingly, for this intermediate range of bubble elongation,  $\Delta\gamma_{\mathrm{max}}$ does not  significantly depend on the bubble orientation with respect to the shear, as shown in Fig. \ref{aspect_ratio_1sur2}. In this graph, we plot $\langle \Delta\gamma_{\mathrm{max}} (\psi) \rangle$ as a function of $\psi$,  obtained by averaging over all values of $\alpha$, $\beta$ and $\theta$ in ${\cal S}$ verifying $\psi$ in the range $[\psi - \delta \psi ; \psi+  \delta \psi]$ with the binning parameter $\delta \psi = 0.5^\circ$. For each value of $\psi$ we also calculate the standard deviation $\sigma_\psi$ on the same subset.
As seen in the figure, the two quantities $\langle \Delta\gamma_{\mathrm{max}} (\psi) \rangle$ and $\sigma_\psi$ only slightly vary with $\psi$. In particular, at any $\psi$ we have $\sigma_\psi \approx \sigma_{\cal S}$, which shows the poor correlation between the bubble orientation and $\Delta\gamma_{\mathrm{max}}$.

\begin{figure}[!ht]
\captionsetup[subfigure]{justification=centering}
\centering
    \begin{subfigure}[b]{0.45\textwidth}
        \includegraphics[width=\textwidth]{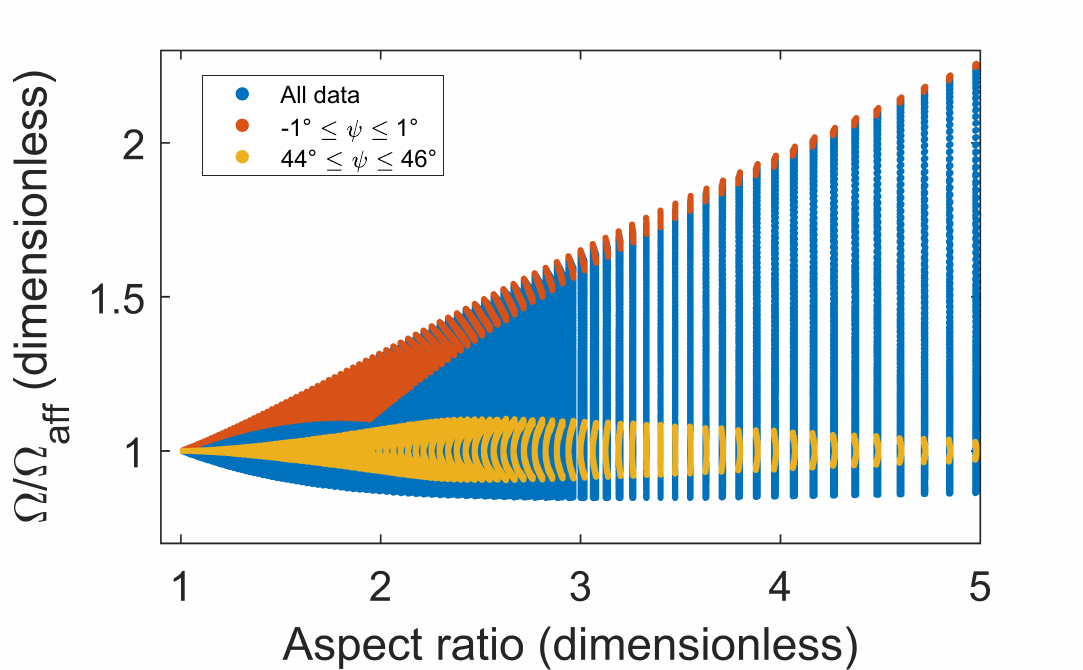}
        \caption{}
    \end{subfigure}
    ~ 
    \begin{subfigure}[b]{0.45\textwidth}
        \includegraphics[width=\textwidth]{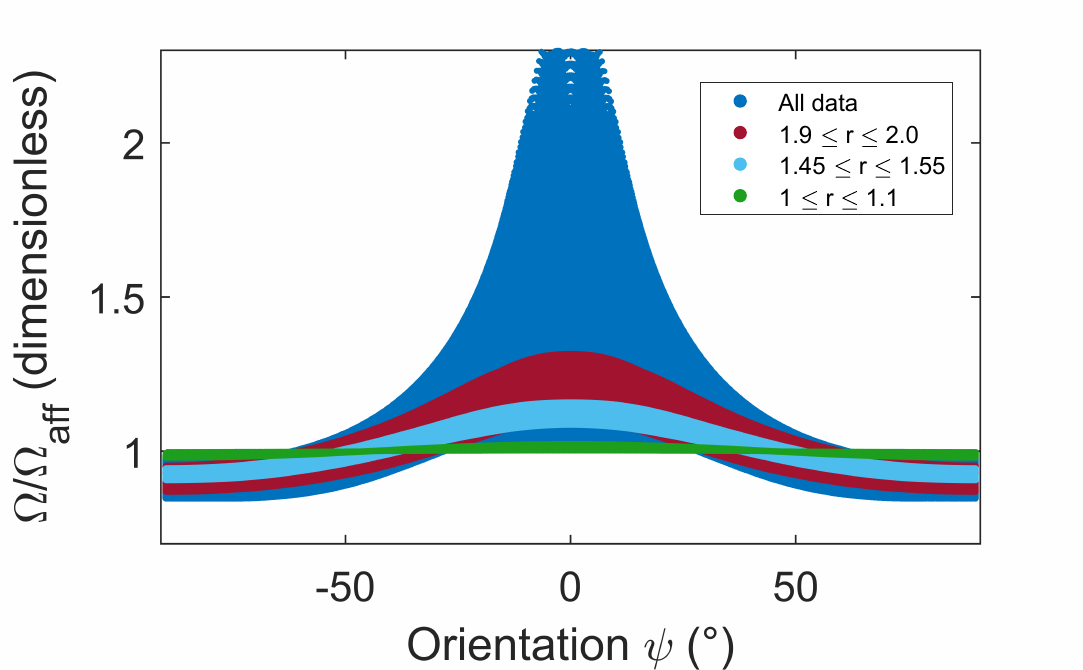}
        \caption{}
    \end{subfigure}
\caption{Color plot of the rotation $\Omega$ rescaled by $\Omega_{aff}$. \textbf{(a)} As a function of the aspect ratio of the bubble $r$ (all points in blue), with highlighted regions corresponding to $-1^\circ \leq \psi \leq 1^\circ$ (orange), and $44^\circ \leq \psi \leq 46^\circ$ (yellow). \textbf{(b)} As a function of the orientation of the bubble $\psi$= (all points in blue), with highlighted regions corresponding to $1.9 \leq r \leq 2$ (burgundy),  $1.45 \leq r \leq 1.55$ (light blue), and  $1 \leq r \leq 1.1$ (green).}\label{omega}
\end{figure}

Finally, in order to quantify the global rotation of the bubble on itself, we compute the total angular momentum of the bubble's perimeter around its center $C^0$. 
\begin{equation}
\Omega = \frac{L}{dt} \int_{s=0}^1  \overrightarrow{C^0 P}(s) \wedge  \overrightarrow{P(s,t+dt) P(s,t)} ds
\end{equation}
This value can be compared with $\Omega_{\mathrm{aff}}$ the total angular momentum of the bubble's perimeter that is be obtained when the external shear $\ep$ is applied globally to the foam structure (in this case, the new position of each vertex is simply computed using eq. \ref{ext_shear}, and the angles between the edges are no longer equal to $120^\circ$). The values of $\Omega/\Omega_{\mathrm{aff}}$ as a function of $r$ and $\psi$ are shown in Fig. \ref{omega}. The range of rotation of the bubble is increased when $r$ is increased, and the biggest rotations are obtained when $\psi = 0^\circ$, i.e. when the long side of bubble is horizontal.

\section{Conclusion}

In conclusion, we have shown, in the case of a 2D periodic foam made of hexagonal bubbles of high Gibbs elasticity, that a global shear applied on the foam necessarily induces a variation of the surface tension along the bubble perimeter, as well as a rotation of the bubble on itself. In this very simple case all the quantities of interest can be analytically computed for any bubble initial shape and any foam deformation. This  model can easily be extended to more complex 2D situations, where the surfactant monolayer is described with a more realistic model than an elastic shell, or for out-of-equilibrium bubble shapes, even though an analytical solution might be out of range in these cases.
Moreover, despite the simplicity of the hypotheses made, the resolution of our model highlights the importance of considering the whole bubble and not simply an isolated  fluid film when dealing with foam deformations.
This crucial question of boundary condition for the surface tension should remain valid in 3D. The  constraint of surface tension continuity we used in this model can be extended to the 2D surface limiting a 3D bubble: the surface tension obtained by integration of the Marangoni law along any closed curve on this surface must come back to its initial value when coming back to the starting point. This seemingly obvious rule can not be addressed on a piece of interface extracted from a larger, closed, interface, so the importance of dealing with the global interface is still pregnant in 3D situations.

\section*{Appendix}

During the shear increment $\Delta \varepsilon^{ext}$, the total length increases from $L_0$ at $t$  to  $L_0+ \Delta L$ at $t+ \Delta t$. Using a Lagrangian formalism, we call $\xi(S_0) = (dS-dS_0)/dS_0$ the local elongation of the interface element $dS_0$ at the curvilinear abscissa $S_0$ (measured at $t$) and decompose it into  $\xi = \Delta L/L_0 + \delta \xi(S_0)$, so that $\int_0^{L_0} \delta \xi(S_0) dS_0 =0$. \\
The local rules are the linearised thermodynamical relationship:
\begin{equation}
\gamma = \gamma_0 +E \xi =\gamma_0 + E \left(\frac{\Delta L}{L_0} + \delta \xi \right) \; , 
\label{thermo1} 
\end{equation}
and the Marangoni law:
\begin{equation}
\frac{\partial \gamma}{\partial S_0} =  \frac{\eta}{h} \frac{\Delta S}{\Delta t} (S_0) \; ,
\label{marangoni}
\end{equation}
with $\Delta S(S_0)$ the distance between the two points that were coincident points at $t$. This term can be decomposed as $\Delta S(S_0) = \Delta_0 S + \delta S$, with $\Delta_0 S$ the value obtained for an homogeneous extension (this part is the one computed in the article) and $\delta S$ the contribution of a non-homogeneous extension $\delta \xi$. 

Noting that the term $\frac{\Delta L}{L_0}$ in eq. \ref{thermo1} is the same at any curvilinear axis, it follows that:
\begin{equation}
\frac{\partial \gamma}{\partial S_0} = E \frac{\partial \delta \xi}{\partial S_0} \; .
\label{thermo2}
\end{equation}
And therefore:
\begin{equation}
E \frac{\partial \delta \xi}{\partial S_0} = \frac{\eta}{h \Delta t}  (\Delta_0 S + \delta S) \; .
\label{thermo_marango}
\end{equation}

The large Gibbs modulus hypothesis established below is based on this last equation: a large Gibbs modulus  implies that variations of $\delta \xi$ are very small, otherwise any stretching or compression of the interface would induce an elastic stress much larger than the internal viscous stress able to occur in the liquid film.

The deformation scale of the problem  is given by $\dot{\varepsilon}^{ext} L$ and  $(\Delta_0 S + \delta S)/\Delta t$ is thus at most of this order. This allows to quantify the order of magnitude of the elongation fluctuations
\begin{equation}
\frac{\partial \delta \xi}{\partial S_0} \sim \frac{\eta}{ E h } \dot{\varepsilon}^{ext} L  \; .
\label{thermo_marango2}
\end{equation}

Using the fact that $\partial \delta \xi/\partial S_0$ scales as $\delta \xi/L$ 
we obtain that $\delta \xi \ll 1$ if 
\begin{equation}
E \gg \frac{\eta L^2}{h} \dot{\varepsilon}^{ext}  \; . 
\end{equation}

In this limit, we can use $\delta \xi$ as a small parameter to make a  Taylor expansion  of the various quantities. Especially, $\delta S$ in eq. \ref{thermo_marango} is of a higher order in $\delta \xi$ than $\Delta_0 S$. At lowest order, the equations \ref{thermo1} and \ref{thermo_marango} then becomes:
\begin{align}
& \gamma = \gamma_0 + E \left(\frac{\Delta L}{L} + \delta \xi \right) \; , \label{27}\\
& E \frac{\partial \delta \xi}{\partial S_0} = \frac{\eta}{h \Delta t}  (\Delta_0 S) \; . \label{28}
\end{align}
This implies:
\begin{equation}
\frac{\partial \gamma}{\partial S_0} = \frac{\eta}{h \Delta t}  (\Delta_0 S) \; ,
\label{29}
\end{equation}
that is the equation we use in the main text.

Note that the Gibbs modulus is large and $1/E$ is of order 1 in $\delta \xi$ (see eq. \ref{thermo_marango2}). Eq. \ref{28} thus consistently compares two terms of order 0 and can be used to compute $\delta \xi$, on the basis of the solutions at dominant order obtained with eq. \ref{29}. 

\section*{Conflicts of interest}
There are no conflicts to declare.

\section*{Acknowledgments}

We acknowledge enlightening discussions with Benjamin Dollet. This project has received funding from the European Research Council (ERC) under the European Union's Horizon 2020 research and innovation programme (grant agreement No 725094). The Matlab code that generates the data presented in this article is available in Zenodo repository: \href{https://doi.org/10.5281/zenodo.1409638}{10.5281/zenodo.1409638}.

\balance


\providecommand*{\mcitethebibliography}{\thebibliography}
\csname @ifundefined\endcsname{endmcitethebibliography}
{\let\endmcitethebibliography\endthebibliography}{}

\end{document}